\documentclass[aps,prl,reprint,superscriptaddress,showpacs,amsmath,amssymb]{revtex4-2}

\usepackage[colorlinks,
           bookmarks=true,
           linkcolor=blue,
           urlcolor=blue, 
            anchorcolor=black,
            citecolor=blue
            ]{hyperref}

\usepackage{multirow}
\usepackage{graphicx}
\usepackage{dcolumn}
\usepackage{bm}
\usepackage{float}
\usepackage{exscale}
\usepackage{relsize}
\usepackage{amsmath}
\usepackage{amssymb}

\begin{document}

\title{The charmonium-like exotic hadron productions in $e^+e^-$ collisions at the BESIII energy with the PACIAE model}
\author{Jian Cao}
\affiliation{School of Physics and Information Technology, Shaanxi Normal University, Xi'an 710119, China}
\author{Wen-Chao Zhang}
\email{Corresponding author: wenchao.zhang@snnu.edu.cn}
\affiliation{School of Physics and Information Technology, Shaanxi Normal University, Xi'an 710119, China}
\author{Zhi-Lei She}     
\affiliation{Wuhan Textile University, Wuhan 430200, China}
\author{An-Ke Lei}
\affiliation{Key Laboratory of Quark and Lepton Physics (MOE) and Institute of
            Particle Physics, Central China Normal University, Wuhan 430079,
            China}
\author{Jin-Peng Zhang}
\affiliation{School of Physics and Information Technology, Shaanxi Normal University, Xi'an 710119, China}
\author{Hua Zheng}
\affiliation{School of Physics and Information Technology, Shaanxi Normal University, Xi'an 710119, China}
\author{Dai-Mei Zhou}
\email{zhoudm@mail.ccnu.edu.cn}
\affiliation{Key Laboratory of Quark and Lepton Physics (MOE) and Institute of
            Particle Physics, Central China Normal University, Wuhan 430079,
            China}
\author{Yu-Liang Yan}
\affiliation{China Institute of Atomic Energy, P. O. Box 275 (10), Beijing
            102413, China}
\author{Zhong-Qi Wang}
\affiliation{Nanjing University of Aeronautics and Astronautics, Nanjing 210000, China}
\author{Ben-Hao Sa}
\email{sabhliuym35@qq.com} 
\affiliation{China Institute of Atomic Energy, P. O. Box 275 (10), Beijing
            102413, China}   
\affiliation{Key Laboratory of Quark and Lepton Physics (MOE) and Institute of
            Particle Physics, Central China Normal University, Wuhan 430079,
            China}
\date{\today}

\begin{abstract}

Inspired by the BESIII observation of exotic hadron $G$(3900) in $e^+e^-\rightarrow D\bar D$ process [PRL 133(2024)081901], we use the parton and hadron cascade model PACIAE to simulate the productions of charmonium-like exotic hadrons including $X(3872)$, $Z_{c}(3900)^0$ and $G(3900)$ in the $e^+e^-$ annihilations at $\sqrt s$=4.95 GeV. The charmonium-like candidates are recombined by Dynamically Constrained Phase-space Coalescence model using component mesons $D\bar D$ (for $G(3900)$) or $D\bar D^*/\bar DD^*$ (for $X(3872)$, $Z_{c}(3900)^0$ and $G(3900)$)  in the PACIAE simulated final hadronic state. We then calculate, for the first time, the charmonium-like candidate's orbital angular momentum quantum number in its rest frame  and  perform the spectral classification for each of the above  candidates. For $G(3900)$,  as its  $J^{PC}$ is $1^{--}$, it is identified as the $P$-wave $D\bar D$ or $D\bar D^*/\bar DD^*$ state. Meanwhile, for $X(3872)$ and $Z_{c}(3900)^0$, as their $J^{PC}$s  are, respectively, $1^{++}$ and $1^{+-}$, they are identified as the $S$-wave $D\bar D^*/\bar DD^*$ candidates. It is observed that the yield of the $P$-wave $G(3900)$ composed of $D\bar D^*/\bar DD^*$ is significantly larger than that composed of $D\bar D$. Moreover, in the $D\bar D^*/\bar DD^*$ channel, the yield of the $P$-wave $G(3900)$ is around three times as large as the yield of the $S$-wave $Z_{c}(3900)^0$, and it is around two orders of magnitude as large as the yield of the $S$-wave $X(3872)$. Finally, significant discrepancies  are observed in the transverse momentum spectra and the rapidity distributions among the $X(3872)$, $Z_{c}(3900)^0$ and $G(3900)$ states. These discrepancies are proposed as valuable criteria for identifying the different charmonium-like exotic hadrons from each other.

\end{abstract}

\maketitle

\begin{bf} Introduction \end{bf} The exotic hadrons (unusual hadrons) 
containing more than three  quarks/antiquarks are allowed and expected 
by the quark model and the Quantum ChromoDynamics 
(QCD) \cite{gell1964,jaffe1977}. The first exotic hadron $X(3872)$ was observed in $e^{+}e^{-}$ collisions by the Belle Collaboration in 2003 \cite{x3872_0}. Since then, more and more experimental measurements and theoretical explanations about the exotic hadrons have been published. There are experimental studies from the BABAR \cite{babar1}, Belle \cite{x3872,belle1,zc3900_1}, BESIII \cite{g_3900,zc3900,zc3900_0, x3872_1}, and LHCb \cite{lhcb1}, etc. collaborations. The 
theoretical methods include the Lattice QCD (LQCD) \cite{chen1,IJMPE2009,gui1,x3872_lattice_1,x3872_lattice_2}, QCD 
sum rule \cite{sumrule_glueball_1,sumrule_x3872_1,sumrule_x3872_2}, the Schr$\rm \ddot{o}$dinger
equation Gaussian expansion method \cite{zhu1}, the quasi three-body unitary model \cite{three_body_model}, and our phenomenological model 
Monte Carlo simulation method~\cite{she2024}, etc..  We refer to the review 
papers in Refs. \cite{chen2016,esposito2017,guo2018,olsen2018,nora2020,chen2023}. 

Our phenomenological model Monte Carlo simulation method composes of the 
parton and hadron cascade model PACIAE~\cite{paciae_3, paciae_4} and the 
quantum statistical mechanics based Dynamically Constrained Phase-space 
Coalescence model (DCPC)~\cite{DCPC}, i.e., PACIAE+DCPC. It has been 
successfully employed in describing the formations of exotic 
hadrons in the $e^+e^-$ annihilations at BESIII energies and in the 
proton-proton (pp) collisions at LHC 
energies~\cite{tai2023,ge2021,hui2022,she2024,zhang2024}. Among our 
method, the model of LQCD \cite{chen1,IJMPE2009,gui1,x3872_lattice_1,x3872_lattice_2}, the QCD sum 
rule \cite{sumrule_glueball_1,sumrule_x3872_1,sumrule_x3872_2}, and the Schr$\rm \ddot{o}$dinger equation  Gaussian expansion method \cite{zhu1}, etc., each has its advantages 
and disadvantages. They are mutual non-exclusion and even indispensable 
mutually. All of them should be studied together with the experiments to 
further reveal the nature of the exotic hadrons. 

After BABAR and Belle collaborations observing a peak around 3.9 GeV in 
$e^+e^-\rightarrow D\bar D/ D\bar D^*$  and 
interpreting it as a $G$(3900) candidate~\cite{babar1,belle1}, the BESIII 
collaboration has also reported its observation of the $G$(3900) peak in the 
$e^+e^-\rightarrow D^0\bar D^0 /D^+D^-$
processes recently~\cite{g_3900}. Meanwhile, its structure has been identified 
as the $P$-wave $D\bar D^*/\bar DD^*$ resonance in Ref. \cite{zhu_2}. The Belle collaboration had measured the mass and the decay width of $X(3872)$ in the decay channel of $X(3872)\rightarrow D^{*0}\bar D^0$ \cite{x3872}. The quantum numbers $J^{PC}$ of $X(3872)$ was determined as $1^{++}$ by the LHCb collaboration \cite{x3872_JPC}. The $Z_{c}(3900)^0$ was firstly observed by the BESIII collaboration in the decay channel of  $Z_{c}(3900)^0 \rightarrow \pi^0 J/\psi$ \cite{zc3900_0}. Later, its $J^{PC}$ was determined as $1^{+-}$ by this collaboration \cite{zc3900_JP,zhu_2}.

In this letter, we use the PACIAE model to simulate the productions of the charmonium-like exotic hadrons including $X(3872)$, $Z_{c}(3900)^0$ and $G(3900)$ in the $e^+e^-$ 
annihilations  at the BESIII upper energy of $\sqrt{s}=$4.95 GeV. The 
charmonium-like  candidate is then recombined by the DCPC model using constituent mesons 
of $D\bar D$  or $D\bar D^*/\bar DD^*$ in the PACIAE model simulated final hadronic state. Here $D\bar D$ refers to $D^0\bar D^0$ or $D^+D^-$, while $D\bar D^*/\bar DD^*$ refers to $D^0\bar D^{*0}$/$\bar D^0 D^{*0}$ or $D^+D^{*-}$/$D^-D^{*+}$. For $G(3900)$, its component mesons could be $D\bar D$ or $D\bar D^*/\bar DD^*$. However, for $X(3872)$ and $Z_{c}(3900)^0$, their component mesons could be only the $D\bar D^*/\bar DD^*$, as they are not observed experimentally in the $D\bar D$ final state \cite{x3872_0,x3872, x3872_1,zc3900_0, zc3900_JP}. We then calculate the charmonium-like candidate's orbital angular momentum  quantum number in its rest frame for the first time. The spectral classification is performed for each of the above charmonium-like candidates 
according to the spectroscopic notation  of $^{2S+1}L_J$ \cite{book1}, where $S$, $L$ and $J$ are, respectively, the spin, the orbital angular momentum, and the total angular momentum (physical spin) quantum numbers of the charmonium-like candidates. The wave shapes of  $S$, $P$, $D$, ... are just corresponding to the $L=$ 0, 1, 2, ...,  respectively.

For $G(3900)$, as its $J^{PC}$ is $1^{--}$ \cite{zhu_2}, it is identified as the $P$-wave $D\bar D$ or $D\bar D^*/\bar DD^*$ state. Moreover, for $X(3872)$ and $Z_{c}(3900)^0$, as their $J^{PC}$s  are, respectively, $1^{++}$ and $1^{+-}$ \cite{zhu_2, x3872_JPC, zc3900_JP}, they are identified as the $S$-wave $D\bar D^*/\bar DD^*$ state. We will compare the yields, the transverse momentum spectra and the rapidity distributions among the $P$-wave $D\bar D$ or $D\bar D^*/\bar DD^*$ state $G(3900)$, the $S$-wave $D\bar D^*/\bar DD^*$ states $X(3872)$ and  $Z_{c}(3900)^0$. These discrepancies will provide valuable criteria for identifying the different charmonium-like exotic hadrons from each other.

\begin{bf} Model \end{bf} 
In PACIAE model, the simulation of $e^+e^-$ annihilation is first executed by PYTHIA6/PYTHIA8 \cite{pythia_6,pythia_8} with presetting of temporarily turning off string fragmentation and with postsetting of breaking up diquark/antidiquark randomly. The partonic initial state is 
available after partons are randomly assigned on the surface of the sphere with radius of 
1 fm. Then this partonic initial state proceeds $2\rightarrow 2$ partonic 
rescattering. Here the Lowest order Perturbative Quantum 
Chromodynamics (pQCD) cross section is introduced for the partonic rescattering processes. For the cases  involving heavy quarks, such as $qc\rightarrow qc$ and $gc\rightarrow gc$, the effect of the heavy quark mass has been considered in the cross section calculation \cite{comb1}. The
final partonic state is reached after the partonic rescattering. Thirdly, the  final partonic state is  hadronized by the LUND string fragmentation regime  and/or Monte Carlo coalescence model. It results an intermediately final 
hadronic state.  Finally, this intermediate final hadronic state goes through the hadronic
rescattering and the kinetic freeze-out happens resulting a final hadronic state (FHS). Here the considered $2\rightarrow 2$ hadronic rescattering 
processes are taken from Ref. \cite{book2} with empirical cross 
section \cite{paciae_3}. 

The DCPC model  is based on the quantum statistical 
mechanics \cite{kobo1965, stowe2007}.  In this model, the yield of $N$-particle cluster reads
\begin{eqnarray}
Y_{N}=\int\cdots\int_{E_\alpha\le E\le E_\beta}\delta_{12\cdots N}\frac{d\boldsymbol{x}_{1}d\boldsymbol{p}_{1}\cdots d\boldsymbol{x}_{N}d\boldsymbol{p}_{N}}{h^{3N}},
\label{eq: two}
\end{eqnarray}
where the $E_{\alpha}$ and $E_{\beta }$ are the cluster's lower and upper
energy thresholds. The $\boldsymbol{x}_{i}$ and $\boldsymbol {p}_{i}$ are, respectively, 
the $i$th particle three-coordinate and three-momentum in the centre-of-mass system (cms) of the $e^+e^-$ collision. It's assumed  that if the 
cluster exists naturally, certain dynamical constraints (the component 
constraint, the coordinate constraint, and the momentum constraint) should be 
satisfied. 

Taking a charmonium-like candidate (i.e., the cluster mentioned above) composed of 
$D^0$ and $\bar D^{*0}$ (i.e., the component hadrons mentioned above) as an example, the dynamical 
constraint is then

\begin{equation}
\delta_{12}=
\begin{cases}
1,& \textrm {if} \ 1\equiv D^0,\ 2\equiv \bar D^{*0},\ R_i\leq  R_0,\ \textrm {and} \atop  m_{\rm 0}-\Delta m \leq m_{\rm inv}\leq m_{\rm 0}+\Delta m   \\ 
0,&  \ \ \ \ \ \ \ \ \ \rm otherwise
\end{cases},\label{eq:delta}
 \end{equation}
where $m_0$ refers to the mass of the charmonium-like candidate, $\Delta m$ is the mass
uncertainty (a free parameter) which is estimated as the half decay width of the charmonium-like candidate.
Table \ref{tab:m_0_and_delta_m} lists $m_0$ and $\Delta m$ for the charmonium-like exotic hadrons $X(3872)$, $Z_{c}(3900)^0$ and $G(3900)$ \cite{x3872,zc3900_JP,g_3900}. $R_0$ and $R_i$ 
($i$=1, 2) are, respectively, the radius of the charmonium-like cluster 
(a free parameter) and the position vector length of the component meson $i$ 
in the rest frame of the cluster. In this frame, the three-coordinate of the 
constituent meson ($D^0$ and $\bar D^{*0}$) is denoted as $\boldsymbol{x}^{*}_i$. It is 
obtained by first Lorentz transforming $\boldsymbol{x}_i$ to the cms of 
constituent mesons and then propagating the earlier freeze-out component meson 
freely to the freeze-out time of the later component meson 
\cite{zhull1,zhull2}. $R_0$ is assumed to be in the range of 1 fm $<R_0<$ 10 fm as the relative distances between two component mesons  should be larger than the sum of the radius of two component mesons and less 
than the interaction range of 20 fm \cite{radius_1, radius_2}. The invariant 
mass ($m_{\rm inv}$) reads
\begin{eqnarray}
m_{\textrm{inv}}=\sqrt{\bigg(\sum^{2}_{i=1} E_i \bigg)^2-\bigg(\sum^{2}_{i=1}
\boldsymbol{p}_i \bigg)^2},
\label{eq:two_3}
\end{eqnarray}
where $E_i$ and $\boldsymbol{p}_i$ ($i$=1, 2) are the component meson's 
($D^0$ and $\bar D^{*0}$ ) energy and three-momentum in the cms of the $e^+e^-$ collisions, respectively.

\begin{table}[]
\caption{$m_0$ and $\Delta m$ for the charmonium-like exotic hadrons $X(3872)$, $Z_{c}(3900)^0$ and $G(3900)$.}\label{tab:m_0_and_delta_m}
	\begin{ruledtabular}
	\begin{tabular}{ccccc}
		   & $X(3872)$ & $Z_{c}(3900)^0$& $G(3900)$ \\
		\colrule
		$m_0$ (MeV/c$^2$)  & 3872.9 & 3893.0& 3872.5 \\
        $\Delta m$ (MeV/c$^2$) & 1.95 & 22.1& 89.85\\
	\end{tabular}
    \end{ruledtabular}
\end{table}

This PACIAE+DCPC model has been successfully applied to 
calculate the yield of the exotic states such as the 
$X$(3872) \cite{ge2021, tai2023, she2024}, $Z_c^{\pm}(3900)$ \cite{zc_3900}, 
$P_c(4312)$, $P_c(4440)$, $P_c(4457)$ \cite{hui2022},  
$\Omega_c^0$ \cite{Omega_c}, and $X$(2370) \cite{zhang2024} based on the PACIAE simulated patonic or hadronic final states.

To generate the charmonium-like candidate composed of $D\bar D$ for instance, a 
component meson list consists of $D^+$,  $D^-$, $D^0$, $\bar D^0$ taken from the PACIAE model simulated FHS should be first constructed. Then a two-layer cycle over the component mesons in the list is then implemented. Each combination in the cycle, if it is $D^+D^-$ or $D^0\bar D^0$ and satisfies the corresponding constraints in Eq. (\ref{eq:delta}), counts as a charmonium-like candidate. The meson list is then updated by removing the used 
meson. A new two-layer cycle is executed on the updated list. Repeat these 
steps until the empty of the list or the rest in the list is unable to 
generate a charmonium-like candidate. The production of the charmonium-like candidate composed of $D\bar D^*/\bar DD^*$ is done in a similar way. 

\begin{figure*}[htbp]
\centering
\includegraphics[scale=0.38]{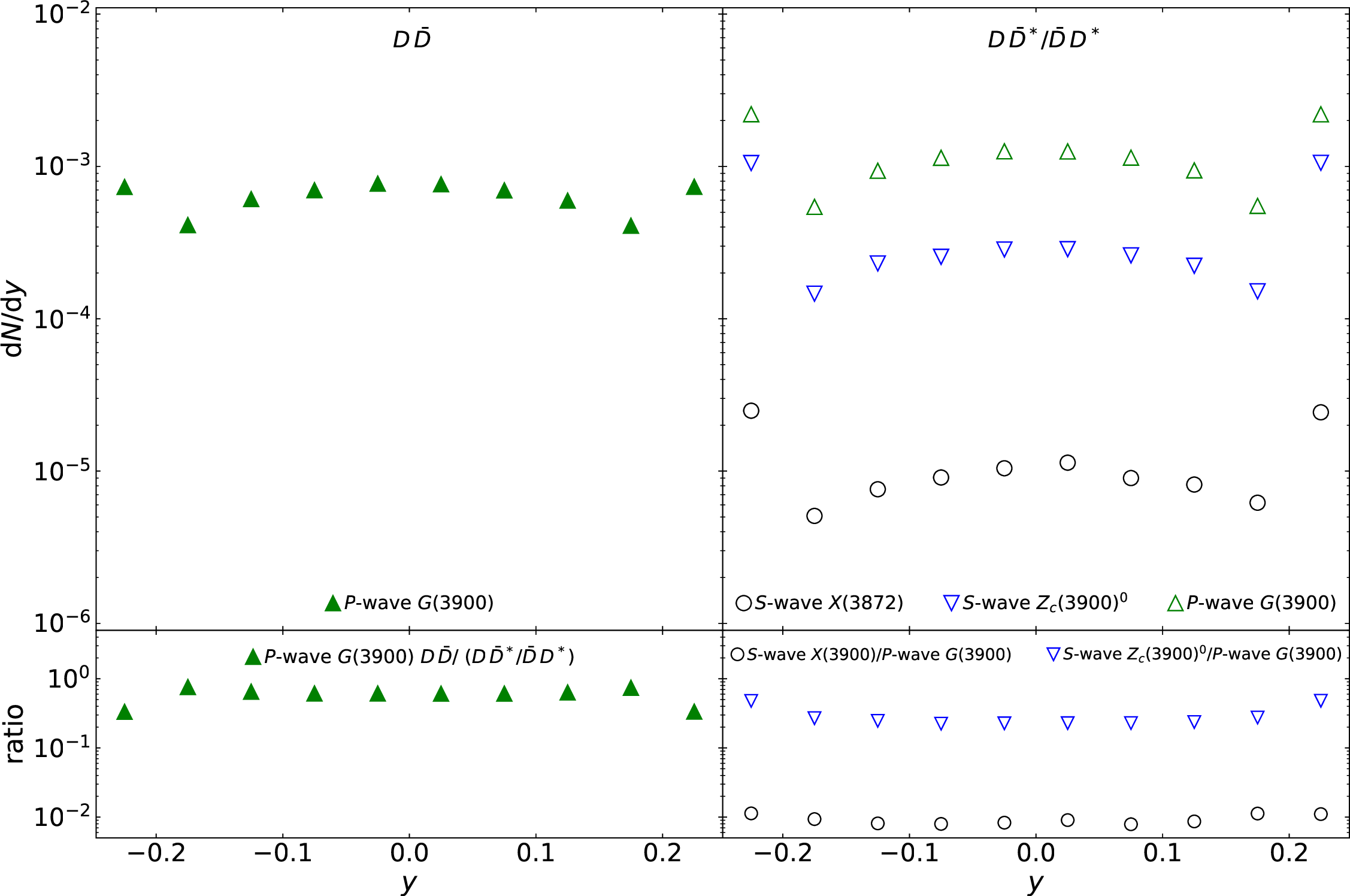}
\caption{\label{fig:yy_charmonium}  Upper panels: the simulated $y$ 
single-differential distributions of the $P$-wave $D\bar D$ (solid triangles-up) and $D\bar D^*/\bar DD^*$ (empty triangles-up) state $G(3900)$, as well as the $S$-wave $D\bar D^*/\bar DD^*$ states $X(3872)$ (empty circles) and $Z_{c}(3900)^0$ (empty triangles-down)  in $e^+e^-$ collisions at $\sqrt{s}$= 4.95 GeV. Lower panels: the ratio between two distributions denoted by legend.}
\end{figure*}


In the charmonium-like candidate rest frame, its orbital angular momentum (OAM) $\boldsymbol{l}^{*}$ is the sum over the OAMs of its constituent mesons. This is expressed as 
\begin{eqnarray}
\boldsymbol{l}^{*}=\boldsymbol{x}_1^{*}\times \boldsymbol{p}^{*}_1+\boldsymbol{x}^{*}_2\times \boldsymbol{p}^{*}_2
\label{eq:OAM}
\end{eqnarray}
where  $\boldsymbol{p}^{*}_i$ ($i$=1, 2) is the three-momentum of the constituent meson. It is obtained  by Lorentz transforming  $\boldsymbol{p}_i$  to the cms of constituent mesons (i.e., the charmonium-like candidate rest frame). According to quantum mechanics, the OAM should be quantized by 
\begin{eqnarray}
\boldsymbol{l}^{*2}=L(L+1)\hbar^2,
\label{eq:OAM_number}
\end{eqnarray}
where  $\hbar$ is the reduced Planck constant and $L$ is  the 
orbital angular momentum quantum number of charmonium-like candidate. As $L$ should be an integer, the solution of Eq. (\ref{eq:OAM_number})
reads
\begin{eqnarray}
L=\textrm{round}\left(\frac{-1+\sqrt{1+4\boldsymbol{l}^{*2}/\hbar^2}}{2}\right),
\label{eq:OAM_method}
\end{eqnarray}
where the function round($X$) returns $X$'s nearest integer. Such a 
quantization rule is denoted as a quantum axiom in Ref. \cite{quantum_axiom}.

\begin{figure*}[htbp]
\centering
\includegraphics[scale=0.38]{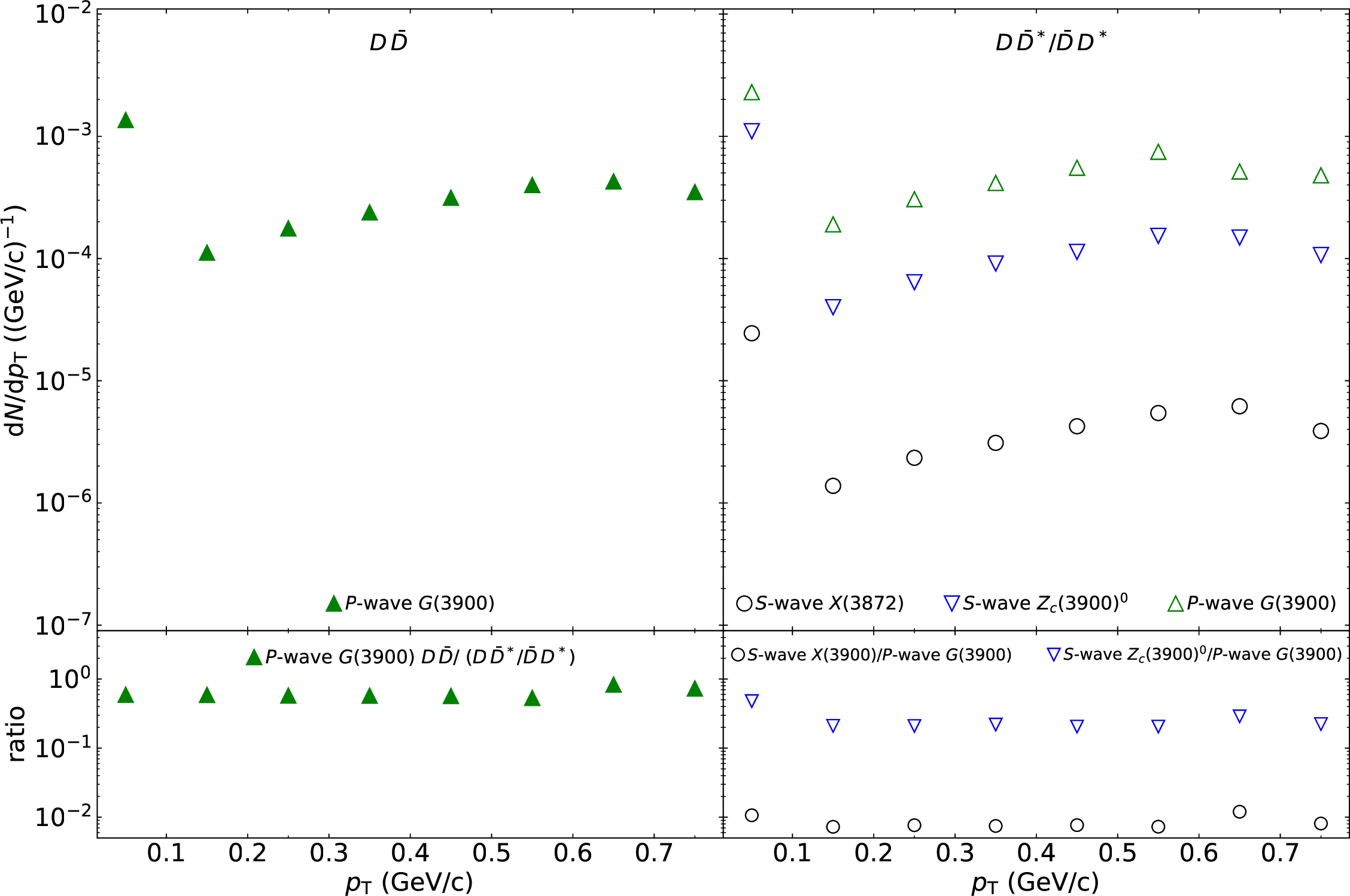}
\caption{\label{fig:pt_charmonium}  Upper panels: the simulated $p_{\rm T}$
single-differential distributions of the $P$-wave $D\bar D$ (solid triangles-up) and $D\bar D^*/\bar DD^*$ (empty triangles-up) state $G(3900)$, as well as the $S$-wave $D\bar D^*/\bar DD^*$ states $X(3872)$ (empty circles) and $Z_{c}(3900)^0$ (empty triangles-down) in $e^+e^-$ collisions at $\sqrt{s}$= 4.95 GeV. Lower panels: the ratio between two distributions denoted by legend.}
\end{figure*}

For the charmonium-like candidate composed of $D\bar D$, its parity is $P=P_D*P_{\bar D}*(-1)^{L}=(-1)^{L}$, and its $C$-parity is $C=(-1)^{L+S}$, where $S$ is the total spin of  $D$ and $\bar D$ \cite{book1}. For the charmonium-like candidate composed of $D\bar D^*/\bar DD^*$, its parity is $P=P_{D(\bar D^*)}*P_{\bar D^*(D^*)}(-1)^{L}=(-1)^{L}$. Since $D\bar D^*$ is not an eigenstate of the charge conjugate operator $C$, the $C$-parity is defined for the symmetric/antisymmetric combination $(D\bar D^*\pm \bar DD^*)/\sqrt{2}$. For the symmetric(antisymmetric) combination, its $C$-parity is $-1(+1)$ \cite{zhu_2}. Having the spin $S$ and the orbital angular momentum quantum number $L$, we are able to determine the charmonium-like state's quantum number $J^{PC}$, where $J=|L-S|,\cdots, L+S$. For the general exotic hadron, its quantum number $J^{PC}$ has to be known by the five-dimensional angular correlation and/or the cross section angular distribution analysis~\cite{x3872_OAM,amplitude_analysis}.




\begin{bf} Results and discussions \end{bf} 
The PACIAE 4.0 model \cite{paciae_4} is used to simulate the $e^+e^-$ collisions at $\sqrt{s}=$4.95 GeV. Total of 500 million $e^+e^-$ collision events are generated with the default model parameters.
Then the charmonium-like  candidates are recombined by the DCPC model with component 
mesons of $D\bar D$  or $D\bar D^*/\bar DD^*$  in the PACAIE simulated FHS. According to the method in the end of the  “Model” section, the $J^{PC}$s for the charmonium-like $S$-, $P$-, and $D$-wave candidates  composed of $D\bar D$ are, respectively, $0^{++}$, $1^{--}$ and $2^{++}$. Moreover, the $J^{PC}$s for the $S$-, $P$-, and $D$-wave  charmonium-like candidates  composed of $D\bar D^*/\bar DD^*$ are, respectively, $1^{+\pm}$, $(0,1,2)^{-\pm}$ and $(1,2,3)^{+\pm}$. For $G(3900)$, as its $J^{PC}$ is $1^{--}$ \cite{zhu_2}, it is identified as the $P$-wave $D\bar D$ or $D\bar D^*/\bar DD^*$ state. Meanwhile, for $X(3872)$ and $Z_{c}(3900)^0$, as their $J^{PC}$s  are, respectively, $1^{++}$ and $1^{+-}$ \cite{x3872_JPC,zc3900_JP, zhu_2}, they are identified as the $S$-wave $D\bar D^*/\bar DD^*$ state. The event averaged yields of the $P$-wave $D\bar D$ and  $D\bar D^*/\bar DD^*$ state $G(3900)$, as well as the $S$-wave $D\bar D^*/\bar DD^*$ states $X(3872)$ and $Z_{c}(3900)^0$  are tabulated in  Table \ref{tab:G_3900_yield}. It is observed that the yield of the $P$-wave $G(3900)$ composed of $D\bar D^*/\bar DD^*$ is significantly larger than that composed of $D\bar D$. Moreover, in the $D\bar D^*/\bar DD^*$ channel, the yield of the $P$-wave $G(3900)$ is around three times as large as the yield of the $S$-wave $Z_{c}(3900)^0$, and it is around two orders of magnitude as large as the yield of the $S$-wave $X(3872)$. 

We have also investigated the effect of varying the coalescence radius $R_0$ on the yield of charmonium-like candidates.  With increasing $R_0$ from 1 fm to 10 fm, the yields of the charmonium-like candidates first increase and then saturate at $R_0\approx$ 2.5 fm. Moreover, in order to study the impact of different methods to extract the value of $L$ for the charmonium-like candidates, another function “trunc($X$)” is utilized in Eq. (\ref{eq:OAM_method}). It truncates the number $X$ to an integer by removing the fractional part of $X$. With the change of the method from “round($X$)” to “trunc($X$)”, the yields of the $P$-wave $D\bar D$ or $D\bar D^*/\bar DD^*$ state $G$(3900) and the $S$-wave $D\bar D^*/\bar DD^*$ state $Z_c(3900)^0$ will change. However, the yield of the $S$-wave $D\bar D^*/\bar DD^*$ state $X(3872)$ keeps the same. Meanwhile, the hierarchy for the yields of the charmonium-like exotic hadrons composed of $D\bar D^*/\bar DD^*$ does not change, i.e. $\textrm{yield}(G(3900))>\textrm{yield}(Z_c(3900)^0)>\textrm{yield}(X(3872))$.



\begin{table}[]
\caption{The PACIAE+DCPC model simulated event averaged yield of the $P$-wave $D\bar D$ and  $D\bar D^*/\bar DD^*$ state $G(3900)$, as well as the $S$-wave $D\bar D^*/\bar DD^*$ states $X(3872)$ and $Z_{c}(3900)^0$ in $e^+e^-$ collisions at $\sqrt{s}$= 4.95 GeV.}\label{tab:G_3900_yield}
\begin{ruledtabular}
\begin{tabular}{ccccc}
 & $P$-wave     & $P$-wave        &   $S$-wave       &  $S$-wave      \\ 
 & $D\bar D$     & $D\bar D^*/\bar DD^*$        &   $D\bar D^*/\bar DD^*$      &  $D\bar D^*/\bar DD^*$      \\
 & $G(3900)$     & $G(3900)$        &   $X(3872)$      &  $Z_c(3900)^0$      \\
 \colrule
yield &  3.87$\times 10^{-4}$     &   6.08$\times 10^{-4}$      & 5.81$\times 10^{-6}$        &    1.97$\times 10^{-4} $        
\end{tabular}
\end{ruledtabular}
\end{table}


The upper panels in Figs. \ref{fig:yy_charmonium} and  \ref{fig:pt_charmonium} show, respectively, the simulated $y$  and $p_{\rm T}$ single-differential distributions for the $P$-wave $D\bar D$ (solid triangles-up) and $D\bar D^*/\bar DD^*$ (empty triangles-up) state $G(3900)$, as well as the $S$-wave $D\bar D^*/\bar DD^*$ states $X(3872)$ (empty circles) and $Z_{c}(3900)^0$ (empty triangles-down) in $e^+e^-$ collisions at $\sqrt{s}$= 4.95 GeV. In the upper panels of Fig. \ref{fig:yy_charmonium}, there is a peak in both the first bin of [-0.25, -0.2] and the last bin of [0.2, 0.25]. Moreover, in the upper panels of Fig. \ref{fig:pt_charmonium}, there is a peak in the first $p_{\rm T}$ bin of 0-0.1 GeV/c. This is because among our simulated final state events of $e^+e^-$ annihilations, aimed on the $D\bar D$ 
production for instance, there are not only two-hadron events of $D+\bar D$ 
but also more than two-hadron events of $D+\bar D+\pi^{+}+\pi^{-}$. In the former events, the transverse momentum of $D$ is just opposite to that of $\bar D$ leading to the zero transverse momentum of the charmonium-like candidate if the final state photons  are neglected, which is responsible for the peaks mentioned in the above rapidity and  transverse momentum distributions. The latter events are responsible for 
the rest of the $p_{\rm T}$ distribution. So far, there is no experimental 
results of the $X(3872)$, $Z_{c}(3900)^0$ and $G(3900)$ productions in the 
$e^+ + e^- \rightarrow \pi^{+} + \pi^{-} + D + \bar D$ and 
$e^+ + e^- \rightarrow \pi^{+} + \pi^{-} + D + \bar D^*$~\cite{private_com}. 
Thus, our results could be served as a prediction of these exotic hadron productions in the $e^+e^-$ annihilation processes with  four-hadron final states containing a pair of charmed mesons.





In both $y$ and $p_{\rm T}$ single-differential distributions, obvious discrepancies are observed among the $P$-wave $D\bar D$ or $D\bar D^*/\bar DD^*$ state $G(3900)$, the $S$-wave $D\bar D^*/\bar DD^*$ states $X(3872)$ and $Z_{c}(3900)^0$. Since the mass of $G$(3900), $X$(3872) and $Z_c(3900)^0$ are very close, it is not easy to distinguish them experimentally.  These discrepancies could serve as effective criteria to identify the different charmonium-like  exotic hadrons from each other. We strongly suggest the experimental measurement of the  $y$  and $p_{\rm T}$ single-differential distributions of the charmonium-like candidates in  the $e^+e^-$ annihilations at the BESIII energies and comparing them with our simulated results.






\begin{acknowledgments}
We would like to thank Shi-Lin Zhu, Xiong-Fei Wang and Lu Meng for the fruitful
discussions. We would also like to thank the anonymous reviewers, whose comments and suggestions are valuable to  improve the quality of the letter. This work is supported by the National Natural Science
Foundation of China under grant Nos. 11447024, 11505108 and 12375135, and by the 111 project of the
foreign expert bureau of China. Y.L.Y. acknowledges the financial support
from Key Laboratory of Quark and Lepton Physics in Central
China Normal University under grant No. QLPL201805 and the Continuous Basic
Scientific Research Project (No, WDJC-2019-13). W.C.Z. is supported
by the Natural Science Basic Research Plan in Shaanxi Province of China
(No. 2023-JCYB-012). H.Z. acknowledges the financial support from
Key Laboratory of Quark and Lepton Physics in Central China Normal University
under grant No. QLPL2024P01.

\end{acknowledgments}


\end{document}